\documentclass[11pt,twocolumn]{article}

\usepackage{graphicx}
\usepackage{float}
\usepackage[bottom=2.5cm, top=2.5cm, left=1.5cm, right=2.5cm]{geometry}

\usepackage{fancyhdr}
\begin{document}

\title{Automated Customization of LLMs for Enterprise Code Repositories Using Semantic Scopes}

\author{
  Ulrich Finkler \\
  email ufinkler@us.ibm.com
  \and 
  Irene Manotas \\
  email irene.manotas@ibm.com 
  \and
  Wei Zhang \\
  email weiz@us.ibm.com
  \and
  Geert Janssen \\
  email geert@us.ibm.com 
  \and
  Octavian Popescu \\
  email o.popescu@us.ibm.com
  \and
  Shyam Ramji \\
  email ramji@us.ibm.com
}

\date{IBM Research\\ Yorktown Heights \\ New York, USA \\[2ex] \today}

\maketitle

\begin{abstract}
Code completion (CC) is a task frequently used by developers when working in collaboration with LLM-based programming assistants. Despite the increased performance of LLMs on public benchmarks, out of the box LLMs still have a hard time generating code that aligns with a private code repository not previously seen by the model's training data. Customizing code LLMs to a private repository provides a way to improve the model performance. In this paper we present our approach for automated LLM customization based on semantic scopes in the code. We evaluate LLMs on real industry cases with two private enterprise code repositories with two customization strategies: Retrieval-Augmented Generation (RAG) and supervised Fine-Tuning (FT). Our mechanism for ingesting the repository's data and formulating the training data pairs with semantic scopes helps models to learn the underlying patterns specific to the repository, providing more precise code to developers and helping to boost their productivity. 
The code completions of moderately sized customized models can be significantly better than those of uncustomized models of much larger capacity. We also include an analysis of customization on two public benchmarks and present opportunities for future work. 
\end{abstract}

\section{Introduction}
\label{sec:intro}

 Generating code snippets is still considered a significant challenge in code intelligence \cite{zan-etal-2023-large, wang-etal-2025-coderag}. Using LLMs for large and well established proprietary code repositories poses additional difficulties. The primary issue is that the code of a proprietary repository has not been seen by even the most advanced LLMs during training. Traditional benchmarks evaluating code completion or code generation as isolated tasks (i.e., not in the context of a repository) can present inflated and misleading results compared to repository-level code tasks \cite{Dewu-humanevo-icse2025}.
 
 Enterprise code often has a specific style, best practice set and customized functionality, even for common tasks as error handling and logging. We worked closely with developers from enterprise repositories to gather key aspects related to the code completion task. 
 Developers identified the 'effort to value' ratio as a key measurement. 
 Having to write a prompt even half the size of the desired code was pointed out as undesirable. 'Near perfection and conciseness' of the prediction was another aspect of high importance. Experienced developers considered fixing a poor prediction a similar amount of work than writing from scratch, just as having to read and trim excessively long predictions. Last, but not least, the latency of the prediction was indicated as highly relevant. Having to wait more than a couple of seconds for a prediction was rated poorly. 

 Although LLMs have shown great progress and performance results for a variety of tasks, more recently Small Language Models (SLMs) have also shown to be sufficiently powerful, inherently more suitable, and necessarily more economical for many invocations \cite{belcak2025smalllanguagemodelsfuture}. Different to what previous work have presented when comparing Retrieval Augmented Generation (RAG) and fine tuning (FT) for code completion \cite{wang2025ragfinetuningcomparativestudy}, we show that fine tuning language models on semantic scopes for code completion is better than RAG when using identical information.

 Based on the developers' input, we focused on a 'minimal effort' scenario to obtain small but high quality predictions as for example filling in the arguments of a function call or adding error handling following the project specific style. The 'minimal effort' consisted of pressing a 'hot-key' with no prompt to generate a prediction that completed a 'code scope' at the current cursor position. Larger predictions can be assembled by triggering the 'hot-key' more than once. 
 The following example illustrates the gaps of even highly rated and large 'out of the box' models when applied for proprietary code. The task is to fill out the arguments for a repo-specific logging call. The desired result (ground truth) is the following:
 \begin{center}
 \begin{scriptsize}
 \begin{verbatim}
            PD_DEV | PD_STACK,
            SQLT_rocmPLFSetPaths,
            0, 0, rc, logPoint,
            PD_LEVEL_SEV,
            PD_3TUPLES_BEGIN,
            PD_STRING( pLeftPath ),
            PD_STRING( pRightPath ),
            PD_UINT( bufLens ),
            PD_STRING( &( dirPath[ 0 ] ) ),
            PD_3TUPLES_END )
\end{verbatim}
\end{scriptsize}
\end{center}
 which includes a tuple of variable arguments that contains symbols from the code prior to the call. 

 Given the entire start of the file up to the desired completion, GPT-OSS-120B  with a specific prompt to complete the function call yields the following result after manually extracting the code snippet from the actual prediction text:
\begin{center}
\begin{scriptsize}
\begin{verbatim}
           SQLT_rocmPLFSetPaths, rc, logPoint,
           PD_STRING( dirPath) )
\end{verbatim}
\end{scriptsize} 
\end{center}
which correctly predicts 3 of the required arguments and one of the optional arguments, but without the necessary 'begin/end' frame. Analogously, GPT-OSS-20B predicts 
\begin{center}
\begin{scriptsize}
\begin{verbatim}
           logPoint, rc, "rocmPLFSetPaths" )
\end{verbatim}
\end{scriptsize} 
\end{center}
which swaps the argument order of the 3 predicted required arguments. Furthermore, this result provides an incorrect location indicator and does not cover the variable arguments. Importantly, even with a situation specific prompt and manual extraction of the code out of the 'wordy' prediction the GPT-OSS models were not able to get close to the style and 'dialect' due to the lack of knowledge about the proprietary repository. 

Even with explicit instructions to only produce the completion code, both models generated large amounts of output surrounding the desired snippet and both models required 30 seconds or more (GPT-OSS-120B up to 100 seconds) on a dedicated H100 GPU with a warmed up model, depending on the specific query.
In comparison, our customized granite-8b-code-instruct model produced in about 1 second the following result: 
\begin{center}
\begin{scriptsize}
\begin{verbatim}
            PD_EXT,
            SQLT_rocmPLFSetPaths,
            0, 0, rc, logPoint,
            PD_LEVEL_ERR,
            PD_3TUPLES_BEGIN,
            PD_STRING( pLeftPath ),
            PD_STRING( pRightPath ),
            PD_UINT( bufLens ),
            PD_3TUPLES_END )
\end{verbatim}
\end{scriptsize} 
\end{center}
predicting usable values for all required arguments and 3 out of the 4 desired optional arguments in the correct style. Importantly, this highlights a general challenge of code completion. Models cannot reliably predict discretionary choices, e.g. the severity level and optional information, unless they are specified in detail in the prompt (which would severely degrade the effort to value ratio). 
Hence, 'guessing' good choices for discretionary values based on the 'style' of a large code base is challenging.

In this paper, we present our method for automated preparation of training data from a large repository such that a fine tuned model generates correct and concise code snippets using the proprietary definitions, naming conventions and coding style of the repository without requiring a user prompt.  Our method is based on a novel approach to create training data pairs based on semantic scopes (see section \ref{sec:approach}) in the code without human labor.

To support on-prem adaptation of models, we customized a language model for two large proprietary repositories containing thousands of files.
Our evaluation on large proprietary repositories shows
that customizing the models via supervised Fine-Tuning (FT) based on semantic scopes achieves the best performance compared to both off-the-shelf (pretrained) models and  Retrieval-Augmented Generation (RAG).
This work makes the following contributions:\\
1) Introducing an automatic repository data ingestion pipeline to prepare training data based on semantic scopes.\\
2) Creating a prototype for customization of LLMs on repository-level data using these training data. \\
3) Presenting our evaluation of the so customized language models and comparing RAG and FT and out-of-the box language models.

This paper is structured as follows. In Section \ref{sec:background} we present related work. In Section \ref{sec:approach} we introduce our approach for repository-level and semantic-scope based data preparation. In Section \ref{sec:methodology}, we describe our customization of LLMs for code tasks.  In Section \ref{sec:results}, we outline the evaluation of our approach on both proprietary enterprise repositories and public benchmarks. Finally, we close by outlining opportunities for future work.

\section{Background}
\label{sec:background}

\begin{figure*}[t]
\centering
  \includegraphics[width=0.8\textwidth]{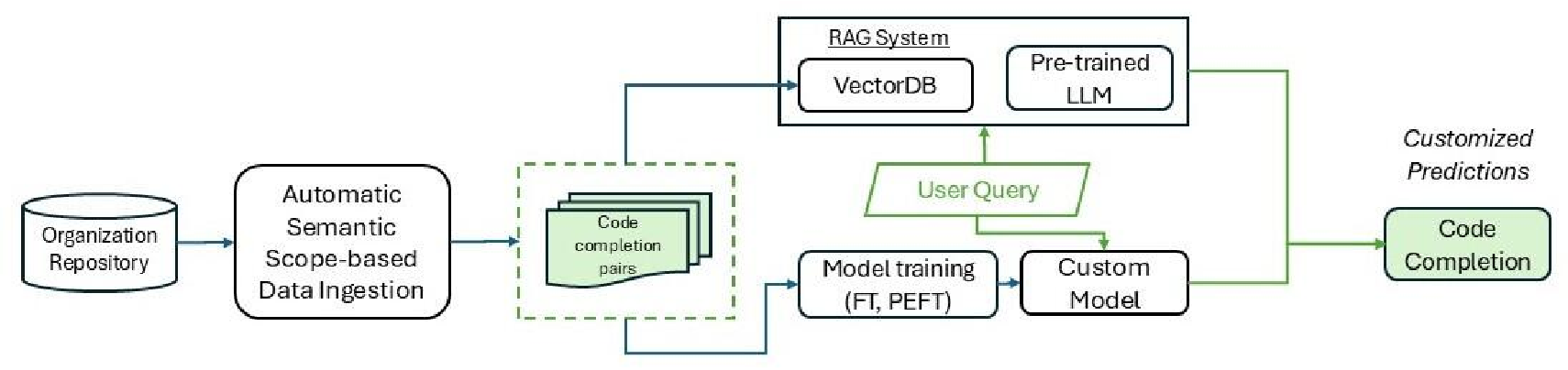}
  \caption{Customization Pipeline Based on Semantic Scopes for Repository-level Code Completion.}
  \label{fig:gralpipeline}
\end{figure*}

\subsection{Repository-level Context for Coding Tasks}
Several benchmarks and previous work have analyzed and proposed techniques to address the LLM code generation task from a natural language prompt without considering the repository context. In such cases, a natural language query describes the purpose of the code to be generated \cite{islam-etal-2024-mapcoder, he-etal-2024-cocost}. However, many of these benchmarks (such as HumanEval) have been identified as not adequately reflecting practical development scenarios \cite{jiang-survey-cg-tosem-2025}.

In contrast, repository-level coding tasks refer to real-world development scenarios in the presence of an existing code base that can provide contextual information and also serve as a knowledge base.

\subsection{LLM Code Completion and Post Training Strategies}

Several approaches have explored how to leverage Retrieval Augmented Generation (RAG) \cite{cheng-etal-2024-dataflow, wang-etal-2025-coderag}, prompting strategies \cite{jiang-survey-cg-tosem-2025, wang2025ragfinetuningcomparativestudy} and post-training (fine-tuning) for the code completion \cite{fim-cc-context-sagtani-wsdm25} and generation tasks. 
DRACO \cite{cheng-etal-2024-dataflow}, a dataflow-guided retrieval augmentation approach improves pre-trained LLMs for the code completion task inside Python repositories. CodeRAG-Bench \cite{wang-etal-2025-coderag} explored RAG for code generation with different retrievers and generators, and found that retrievers often struggle to fetch useful contexts, and generators face limitations in using those contexts effectively. We explore RAG and Fine Tuning (FT) for customizing models on the code completion task based on semantic scopes from the code. We evaluate these two customization strategies on two large proprietary repositories covering Java and C/C++ code.


 Post-training LLMs using a curriculum dataset approach is proposed in \cite{fim-cc-context-sagtani-wsdm25} by extracting hard-to-complete patterns from code repositories and generating context examples using semantic and static analysis tools. The work shows that all fine-tuned models improve for code completion. Performance gains are more pronounced for smaller parameter models.
Wang et. al. \cite{wang2025ragfinetuningcomparativestudy} conduct training at the file level
using source code for customizing code LLMs for the code completion task \cite{wang2025ragfinetuningcomparativestudy}. Different to previous work, we study and show how post-training of LLMs for code completion at the repository-level, on two large and proprietary enterprise repositories for two different languages (Java and C++), perform and compare to a RAG customization approach.

\subsection{Benchmarks}
Several repository-level benchmarks for coding tasks have emerged in the last couple of years \cite{liu2023repobench, ding-cceval-2023, evoCodeBench-NIPS-2024, Dewu-humanevo-icse2025}.
RepoBench-C\cite{liu2023repobench} is a benchmark specifically designed for evaluating repository-level code auto-completion systems with evaluation for Python and Java; CrossCodeEval (CCEval)  \cite{ding-cceval-2023} is a benchmark for code completion based on a diverse set of real-world, open-sourced, permissively-licensed repositories, including a Java dataset and three additional datasets for Python, TypeScript, and C\#. 
\textit{HumanEvo} \cite{Dewu-humanevo-icse2025} is an evolution-aware repository-level code generation dataset and an automated execution-based evaluation tool with 400 samples split evenly between Python and Java.  EvoCodeBench \cite{evoCodeBench-NIPS-2024} is another benchmark with an automatic collection pipeline, which constructs new versions from the latest version of the considered repositories. The size of the dataset is 275 samples collected from 25 repositories. 

Although these benchmarks can provide insights about how a pre-trained LLM performs for different coding tasks, they do not provide a training set to investigate how a tuned model compares to other non-tuning strategies. Therefore, we leverage repository-level code benchmarks having not only test samples but also data for post-training, such as RepoBench \cite{liu2023repobench} and CrossCodeEval (CCEval) \cite{ding-cceval-2023}, to evaluate the tuning of LLMs for code completion and code generation.

\begin{figure*}[t]
\centering
  \includegraphics[width=0.8\textwidth]{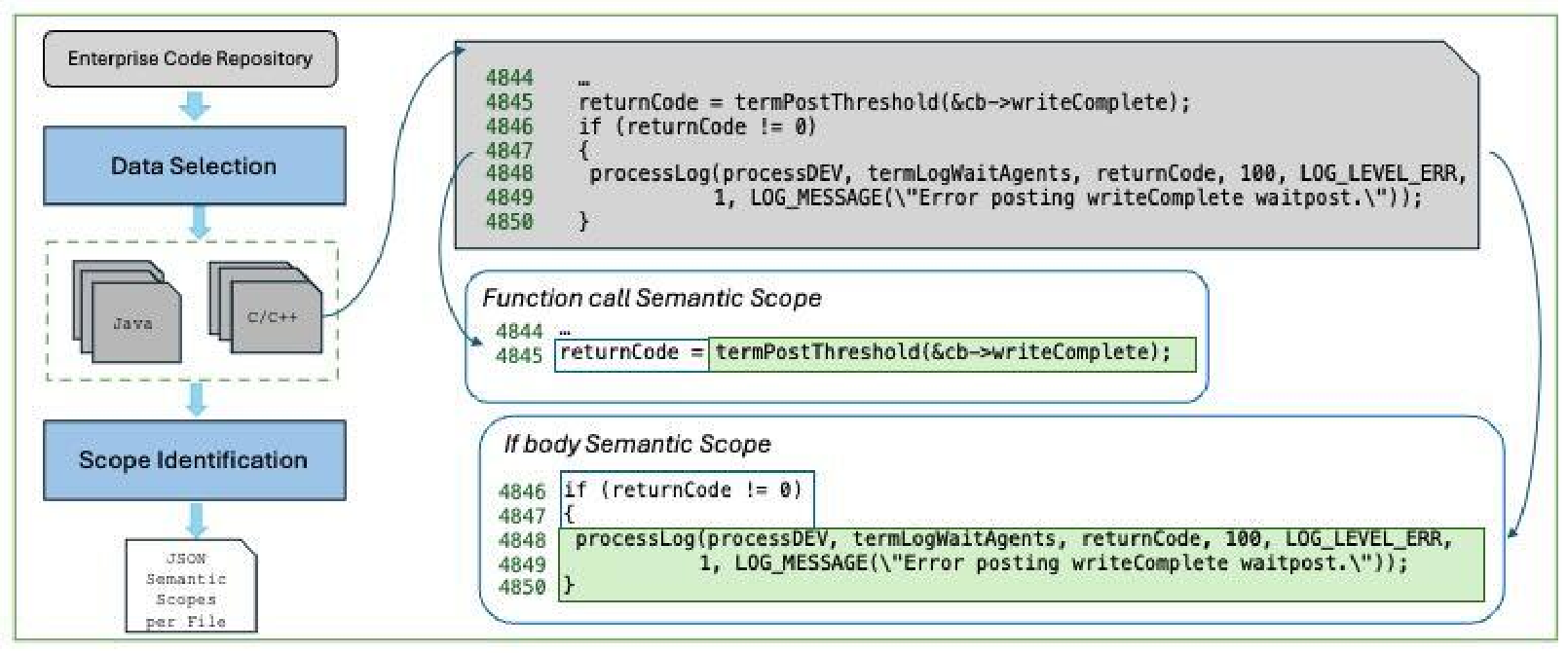}
  \caption{Semantic Scope based Repository Data Ingestion. Left side shows the process flow; Right side shows two examples of semantic scopes (shaded boxes). } 
  \label{fig:semanticscopes}
\end{figure*}

\section{Automated Data Preparation and Model Customization}
\label{sec:approach}

Large industry software projects often have their own coding style and 'dialect', i.e., there is a lower level collection of
proprietary software artifacts. Higher level software artifacts are built using those lower level artifacts.
When working with AI tools, two questions arise:\\
(1) How does one automatically extract training data from a large software repository (often many GB) to teach a model the coding style and dialect?\\
(2) How to teach a model to `guess' how much code completion should a given LLM generate without requiring a prompt?

Figure \ref{fig:gralpipeline} shows the general automated customization pipeline. We ingest the repository data and create sample pairs for code completion (and potentially other coding tasks) based on {\em semantic scopes} of a program. In this section we describe how our language model customization approach works for the code completion task.

\subsection{Automated Data Preparation}

In formal semantics, which focuses on the 'meaning' of text elements and their relationships, a semantic scope is a collection of related text elements, for example parts of text to which a semantic operator like negation may apply. Analogously, code has semantic scopes that are independent of the syntax or programming language. Since semantic scopes are based on meaning, they remain intact even when compiling for example C++ code to assembler, which flattens the syntactic structure to a large degree. Figure \ref{fig:semanticscopes} shows two examples of the semantic scope identification from a C/C++ code snippet.

Semantic scopes provide language independent units to generate training data. Pairing a code prefix with a code snippet that forms a semantic scope teaches the model to generate code completion snippets with a  'meaning' and with proper end-of-text tags. In this way, the generated code snippets complete  the current semantic scope to achieve meaningful conciseness without requiring a specification from the developer. 

Extracting semantic scopes can be challenging and may require data and control flow analysis. But in modern programming languages with functions, classes and different types of clauses, semantic scopes often coincide with syntactic scopes. The enterprise repositories we considered used Java and C++. For these languages, the 'bodies' between matching brackets and parentheses are usable scope candidates as are syntactic constructs found in the syntax tree, e.g. function and method definitions or the bodies of loops and conditionals.

Semantic scopes are recursive, i.e., a larger semantic scope as 'code to load data and compute X' contains a scope loading data and another to 'compute X'. Hence, a repository yields scope candidates of widely varying sizes and compositions that need to be filtered. Filter options for semantic scopes can be the {\em size} of the scope, the {\em depth} (e.g. how many levels of nested brackets are inside a scope), edit time stamps, and keywords (e.g. deprecated symbols). For a given repository and language, a search for good filter parameters can yield useful improvements in model performance. 

In our experiments scopes between at least \(50\) and at most \(1,000\) bytes yielded good results. Another useful filter option is the amount of code available that precedes a scope. A minimum of \(200\) bytes has proven useful in our experiments.

 Our goal is to teach the model to complete the currently open semantic scope. For example if the input code ends with the opening parenthesis of a {\em method invocation}, the model
should complete the {\em argument list}. If the input code ends with a position inside a method's definition, the model should complete the method's definition.
To ingest and process data from a repository for customization our pipeline includes the processes described below.\\
{\em Data Selection:} In this step the pipeline collects all code files from the repository and performs an isolation process where the files are selected by their programming language. Data from each file in the repository is stored in a JSON object with metadata including the path to the file, the date, etc..\\
{\em Scope Identification:} 
 After selecting the code files in a repository, we process each file to extract scope candidates and apply the chosen filters. \\
{\em Generating Completion Pairs:}
The generation of code completion pairs begins with the continuous code sections containing the `prefix' and the `scope' identified as described above. These sequences are partitioned into pairs of a `query' and a `label'. A primary partitioning location is the start byte of the scope, yielding \textit{`primary pairs'}. 
Additional pairs, with a start at a random number of bytes behind the primary location for each candidate, can increase the robustness of the finetuned model. Each `label' is terminated with an `end-of-text' token.

\subsection{LLM Customization}
To improve the performance of LLMs for a given repository, we explore two strategies: Retrieval Augmented Generation (RAG) and  Fine Tuning. 

\subsubsection{RAG}
The primary pairs are the basis for the RAG knowledge base. The {\em queries} are embedded into a high dimensional vector space with a suitable model, for example from the sentence transformers collection. 
The embedding of the query forms the search key. The code of the corresponding label forms the value. 
These key-value pairs are loaded into a suitable vector database. An inference query is embedded and its top-N nearest neighbors are identified. The values of these nearest neighbors are 
used to augment the inference query. 

RAG introduces sources of errors. The vector embedding is an approximation for the correlation between 
queries, i.e. code prefixes. Hence the top-N nearest neighbors may not include the best solution candidates. 
Furthermore, even similar prefixes may use different symbols and thus lead to the selection of symbols 
not occurring in the query prefix when generating the code snippet. Last, but not least, RAG has very 
limited opportunity to teach a model to stop well by only being able to provide a number of examples.

\subsubsection{Fine Tuning}

The primary pairs are also the basis of fine tuning. Since the `prefixes' can be larger than the 
`label', for example we allow up to 3kB of code as a prefix, the masking of the query in the loss 
calculation as well as the addition of an end-of-text token is essential to obtain concise models. 
The `random start' data pairs can be used to augment the primary data for training. 
It increases the robustness of the prediction if the cursor is not at the beginning of 
a scope.

\section{Methodology}
\label{sec:methodology}

In this section we outline our methodology to conduct experiments on enterprise repository data and public benchmarks.

\subsection{Models}

Smaller models are gaining attention due to their ease of deployment and training \cite{chen-kd-fsejournal-2025} and low latency, amongst other reasons. This is particularly true for on-prem deployment. 
Hence, we selected two small models for customization, Llama-3.1-8B-instruct (Llama) and Granite-8B-code-instruct (Granite). To compare with bigger models, we include as a baseline model Qwen2.5-72B.

\subsection{Datasets}

\begin{table}[t!]
\caption{CC Datasets (number of sample pairs)}
\begin{center}
  \scriptsize
\begin{tabular}{|l|l|r|r|r|}
\hline
\textbf{ }&\textbf{Lang.}&\multicolumn{2}{|c|}{\textbf{Splits}} & \textbf{Total}\\
\cline{3-4} 
 \textbf{}   &  & \textbf{Train}  & \textbf{Test} & \\ \hline
DataB (prop) & C/C++ & 200,000 & 173 & 200,173  \\ \hline
STM (prop) & Java & 316,609 & 156 & 316,765  \\ \hline \hline 
RepoBench (pub) & Java & 6,956  & 1,740  & 8,696 \\ \hline
CCEval  (pub) & Java & 1,711 & 428 & 2,139 \\ \hline
\end{tabular}
\end{center}
\label{pri_repos_cc}
\end{table}

We evaluated the performance of our customization approach on two large enterprise code repositories: DataB and STM. The summary of the datasets ingested from the private enterprise repositories are shown in Table \ref{pri_repos_cc}. Here we present the general information of these repositories. 

\textbf{DataB}: A large Database management application repository with around \(27\)K files covering \(13\) million lines of C/C++ code.\\
\textbf{STM}:  A large enterprise cloud-based platform for asset and facilities lifecycle management with around \(13\)K files covering \(1.3\) million lines of Java code. \\

Table \ref{tbl_test_group_summary} summarizes the test case numbers by test group that were manually curated from each enterprise repository. Test data were created manually in close collaboration with the developers from a set of hold out files, 
simulating the scenario of a developer employing code completion while writing new code. Developers directed the types of scenarios and checked the quality. Additionally, 
a search was performed to ensure that the desired prediction was not present in the training data. 

Excluding the entire file from which a test sample was drawn proved essential to achieve decent correlation between performance measurements and user experience. 
The files in the repositories tended to be relatively large and the similarities within a single file tended to exceed those across files. Hence, including parts of files from which tests were drawn in the training data yielded overly positive measurements compared to the experience of a real user requesting completions in new code.


\begin{table}[t!]
\caption{Test Cases by Semantic Scope Categories for STM and DataB Enterprise Repositories Summary}
\begin{center}
  \scriptsize
\begin{tabular}{|l|r|r|}
\hline
        \textbf{Category}        & \multicolumn{2}{c|}{\bf{\# Test Cases}}  \\ \hline
        & \textbf{DataB } & 
        \textbf{STM  }  \\ \hline 
else\_body             & 13  & 20\\ \hline
for\_body             & 14  & 16 \\ \hline
func\_body            & 16  & 38 \\ \hline
if\_body              & 14 & 30  \\ \hline
logging               & 36  & 19 \\ \hline
func\_call    &  80  & 33 \\ \hline

\end{tabular}
\end{center}
\label{tbl_test_group_summary}
\end{table}

Besides the above large enterprise repositories, we selected two public repository-level benchmarks to evaluate the performance of our customization approach: CCEval and RepoBench.\\
\textbf{CrossCodeEval (CCEval)}~\cite{ding-cceval-2023}. A Code Completion (CC) benchmark for single line completion. The goal is to evaluate in-depth cross-file contextual understanding to complete the code accurately. \\
\textbf{RepoBench}~\cite{liu2023repobench}. Samples functions from real-world projects and evaluates how LLMs generate these functions according to the target function description and project context.\\

 Comparing the statistics in table \ref{pri_repos_cc} shows the large size difference between enterprise and public repositories. 

\begin{figure*}
\centering
\includegraphics[width=0.49\textwidth]{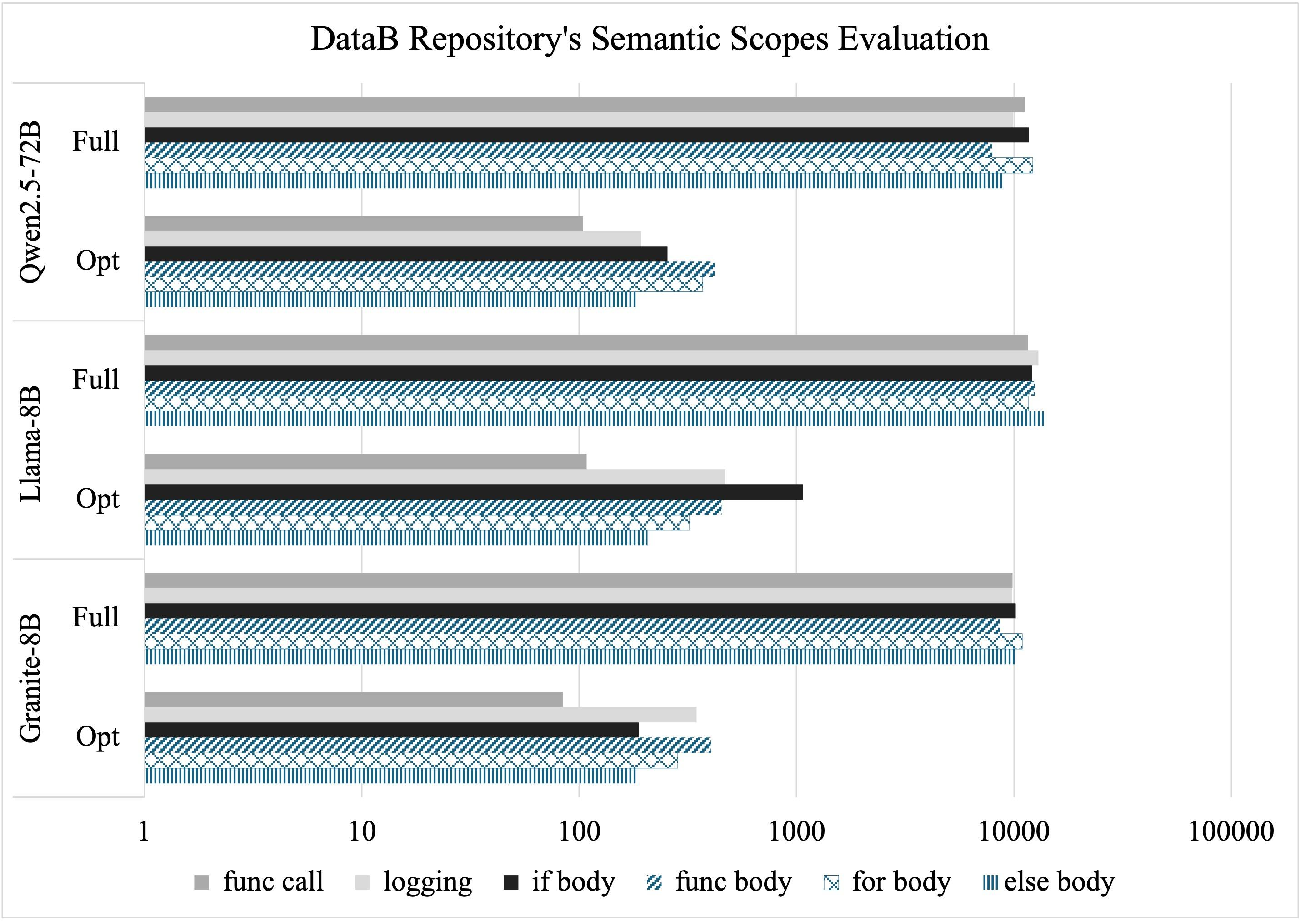}%
\vspace{0.1pt}
\includegraphics[width=0.49\textwidth]{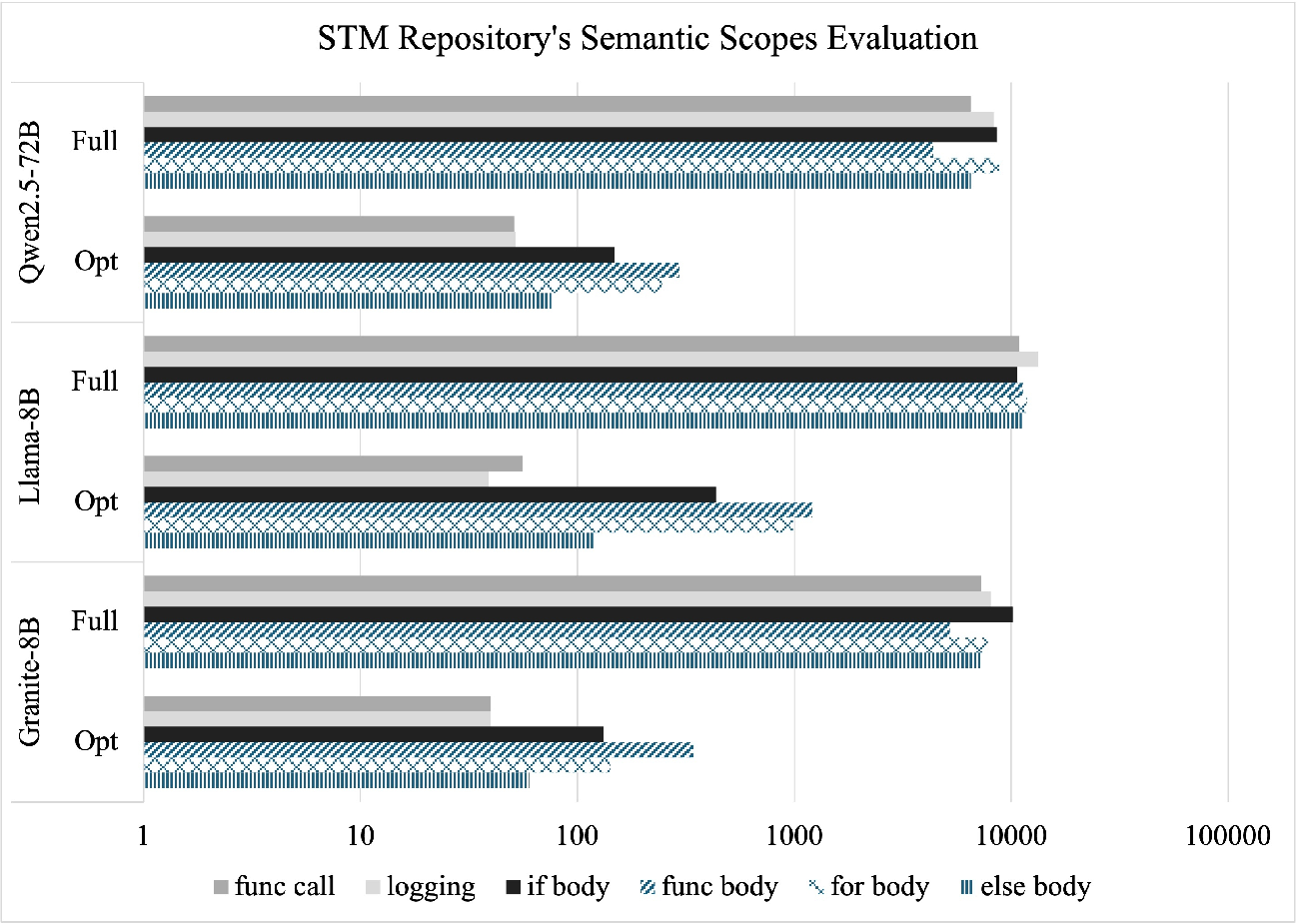}%
\caption{Performance (optimal and full Levenshtein Distance) of Baseline Models shown in logarithmic scale. (Left) Results on DataB repository's test set. (Right) Results on STM repository's test set. Shorter bar is better.}\label{label}
\label{fig:baselines-proprepos}
\end{figure*}

\subsection{Evaluation Metrics}
\label{subsec:metrics}

The {\it Levenshtein distance} \cite{Levenshtein} is a well established distance metric for text comparison and is directly correlated to the effort to correct a prediction into the desired solution. Levenshtein distance also satisfies the requirements of a mathematical metric as for example the Euclidian distance. The properties of a mathematical metric are the foundation for reasoning with measurements in most fields of engineering. 
To compare the generated code with the reference label, a.k.a. ground truth, we considered two types of Levenshtein distances: full and optimal.

The full Levenshtein distance (\textit{Full}) between a prediction and the desired solution helps to understand how close a part of the prediction to the desired answer and how concise the prediction is.
The \textit{Full} metric is higher if a prediction is very different from the desired ground truth, and it is higher when the prediction is excessively long relative to the desired solution. 

The optimal prediction prefix (\textit{Opt}) is the beginning of the prediction, varying the length, that achieves the lowest Levenshtein distance relative to the desired ground truth. This measurement is low if the beginning of the prediction is close to the desired solution, but is not degraded by additional content.

The difference between \textit{Full}  and \textit{Opt} reflects an approximation of how much undesired additional content is predicted. For a {\em concise} prediction, the difference between \textit{Full} and \textit{Opt} is close to zero. For a nearly exact match, the \textit{Opt} distance is also close to zero.

Some popular metrics, e.g. BLEU score or executability, have significant disadvantages in the code completion scenario. 
In code, order is highly relevant. Swapping two arguments in a function call may render it already incorrect. 
The BLEU score (and many other scores for text) do not take order sufficiently into account \cite{policeandrobbers}. Scores limited to a range, e.g. 0 to 1,  in general do not satisfy the properties of a mathematical metric. 

Measurements that count exact match or require parsing or compiling may assign the same result, fail, to a random string and a prediction that requires only the alteration of a single character to match the desired outcome. 
The measurement is far from proportional to developer effort to check and fix the solution. Furthermore, 
the code in 'code completion' is by definition incomplete and may not parse even for a perfect prediction. 

\subsection{Model Customization}
\label{sec:cus-strategy}

Given a set $S$ of 'prefix-scope' pairs (\(200,000\) for DataB and \(260000\) for STM), we evaluated two customization strategies, RAG and supervised fine tuning (FT) utilizing $S$ for the knowledge base and training data, respectively.

\subsubsection{RAG} 

To embed code sections, we used \verb@all-MiniLM-L6-v2@ from the sentence transformer python package. To avoid errors due to approximation algorithms in top-N nearest-neighbor searches we used a linear cost exact search based on cosine distance. A small number of neighbors (3 or 5) as query augmentation preceding the code yielded the best results. 

\subsubsection{FT} 

Our FT training pipeline used the Huggingface Trainer class and infrastructure with an Adam optimizer. One of the major hurdles was numerical instability. Relatively small changes in hyperparameters lead to unexpectedly large variations in model performance. A change in version of the underlying model lead in some cases to significant degradation, despite preserving the number of parameters, the underlying transformer stack and overall architecture to a large degree and despite of reaching a similar final training loss.

Careful investigation of the progression of training loss, gradient norm and test evaluation metrics of intermediate checkpoints revealed that numerical instability was a major contributor. Literature suggests that the Adam Optimizer can be a source of numerical instability \cite{AdamInstability}\cite{AdamInstability2}.
Indeed, offloading the optimizer to CPU, where it is executed in 32 bit floating point arithmetic,
reduced the frequency and height of gradient norm spikes. Avoiding padding and packing also improves the stability of convergence.

Once gradient norm spikes are controlled, test metrics tend to show a 'saturation level' that can be reached via different training settings and sometimes even via different base models, suggesting that the information in the training data dominates the outcome. 

In addition to the Levenshtein metrics, we gathered developer feedback that tested the completion capability in their work flows.

\section{Results}
\label{sec:results}

The most impoartant evaluation of our customizations was the feedback of developers that tested it 
in their normal work environment. Here are couple of testimonies: \\
“I would be interested in using [custom model] because the code suggestion there were more accurate. Suggestions from [custom model] were concise and easy to modify if required. Results for [custom model] were consistent” \\
“Yes, the custom model, although absolutely not tailored towards my use cases outside of DataB code, performed surprisingly well and much better than the [uncustomized] version - it outmatched the other in every single test or was at least on par in general tinkering outside of provided samples.” 

The uncustomized version used in the user study employed the Granite-8B model. The custom model was an FT-customized version (as described above) of that Granite-8B model. In general, users expressed a preference for the customized model for code completion with respect to conciseness and quality.

\subsection{Measurements}

Additionally, we measured the performance of baseline models and customized models for both proprietary repositories and for the public benchmarks in terms of the metrics described in Section \ref{subsec:metrics}. 

The GPT-OSS models are reasoning chat models and thus not suited for the code completion task without a situation specific prompt, making a fair comparison in the 'minimum effort' scenario infeasible.
Figure \ref{fig:baselines-proprepos} shows results for Qwen, Llama and Granite, which were able to generate an at least decent start of a prediction without a situation specific prompt. 
All the 'out of the box' models generated excessive amounts of code without a situation specific prompt. For the 'out of the box' models, the smaller Llama-8B and Granite-8B proved to be competitive for the code completion task on enterprise data across the test categories, even relative to Qwen2.5-72B.  Hence, Llama-8b and Granite-8B will be our references when investigating the benefit of customization. 

Tables \ref{tbl-rag-ft-datab} and \ref{tbl-rag-ft-stm} show the improvements of RAG and FT over the baseline Llama-8b and Granite-8B models. The measurement that correlated best with human evaluations was the full Levenshtein distance, which vastly improved with fine tuning in all cases.  For the 'opt' distance, the improvements vary depending on test category and repository. 

 Note that the metrics are distance measurements (lower is better).
Due to the 'discretionary options' in predictions, a perfect match (distance zero) is extremely unlikely and hence 
there is a 'lower bound' for what the models can realistically achieve. Human inspection reveals that 
a change in the 'opt' Levenshtein distance by 10 edits is significant for measurements in the vicinity of 100, for example from predicting the correct number 
of required arguments with correct types to having one or two arguments out of order, with incorrect type or missing when filling out an argument list as in the example in the introduction. 

\begin{table}[H]
\caption{ Baseline, RAG, and FT for DataB.}
\begin{center}
  \scriptsize
\begin{tabular}{|l|rr|rr|rr|rr|}
\hline
  \textbf{DataB}      & \multicolumn{2}{c|}{\textbf{Granite-8b}} & \multicolumn{2}{c|}{\textbf{Granite-8b}}   & \multicolumn{2}{c|}{ \bf{Granite-8b} }  \\
  \textbf{Tests} & \multicolumn{2}{c|}{\textbf{Base}}    & \multicolumn{2}{c|}{\textbf{RAG}}      & \multicolumn{2}{c|}{ \bf{FT} }  \\ \cline{2-7} 
                      & \multicolumn{1}{r|}{\textbf{Opt}}  & \textbf{Full} & \multicolumn{1}{r|}{\textbf{Opt}}  & \textbf{Full} & \multicolumn{1}{r|}{\textbf{Opt}}  & \textbf{Full}\\ \hline
else\_body         & \multicolumn{1}{r|}{186}  & 10,260  & \multicolumn{1}{r|}{\bf{171}}  & 15,138   & \multicolumn{1}{r|}{189} & \bf{189}  \\ 
\hline
for\_body          & \multicolumn{1}{r|}{284}  & 10,942 & \multicolumn{1}{r|}{301}  &  10,627  & \multicolumn{1}{r|}{\bf{238}} & \bf{815} \\ 
\hline
func\_body        & \multicolumn{1}{r|}{404}  & 8,631 & \multicolumn{1}{r|}{338}  & 9,359    & \multicolumn{1}{r|}{\bf{321}} & \bf{1,266}   \\ 
\hline
if\_body        & \multicolumn{1}{r|}{\bf{188}}  & 10,174    & \multicolumn{1}{r|}{194}  &   10,897  &  \multicolumn{1}{r|}{229} & \bf{239}  \\ 
\hline
logging          & \multicolumn{1}{r|}{347}  &  9,783 & \multicolumn{1}{r|}{170}  & 10,188   & \multicolumn{1}{r|}{\bf{129}}  &  \bf{144}  \\ 
\hline
func\_call         & \multicolumn{1}{r|}{84}  &  9,834  & \multicolumn{1}{r|}{82}  & 9,560   &  \multicolumn{1}{r|}{\bf{42}}   & \bf{45}  \\ 
\hline
\hline
    \textbf{DataB}  &  
  \multicolumn{2}{c|}{ \bf{Llama-8b} } &  \multicolumn{2}{c|}{\textbf{Llama-8b}}  & \multicolumn{2}{c|}{ \bf{Llama-8b} } \\ 
 \textbf{Tests} &  
  \multicolumn{2}{c|}{ \bf{Base} } &
  \multicolumn{2}{c|}{\textbf{RAG}}  & \multicolumn{2}{c|}{ \bf{FT} } \\\cline{2-7} 
                       & \multicolumn{1}{r|}{\textbf{Opt}}  & \textbf{Full} & \multicolumn{1}{r|}{\textbf{Opt}}  & \textbf{Full} & \multicolumn{1}{r|}{\textbf{Opt}}  & \textbf{Full}\\ \hline
else\_body         &  \multicolumn{1}{r|}{206} & 13,741 &  \multicolumn{1}{r|}{222}  &  13,605   & \multicolumn{1}{r|}{\bf{194}} & \bf{203}\\ 
\hline
for\_body         & \multicolumn{1}{r|}{322}  & 11,716  &  \multicolumn{1}{r|}{\bf{309}}  &  11,672  &  \multicolumn{1}{r|}{315} &  \bf{326}\\ 
\hline
func\_body     &  \multicolumn{1}{r|}{452} & 12,462  & \multicolumn{1}{r|}{450}  & 12,024  &  \multicolumn{1}{r|}{\bf{363}} & \bf{368}  \\ 
\hline
if\_body       &    \multicolumn{1}{r|}{1,070} & 12,135  &  \multicolumn{1}{r|}{\bf{248}}  &    10,760  &  \multicolumn{1}{r|}{256} & \bf{280} \\ 
\hline
logging & \multicolumn{1}{r|}{470}  & 12,954 &  \multicolumn{1}{r|}{400}  &    12,309  &  \multicolumn{1}{r|}{\bf{157}}   & \bf{172} \\ 
\hline
func\_call   &  \multicolumn{1}{r|}{108} & 11,592 &  \multicolumn{1}{r|}{101}  &    11,083 & \multicolumn{1}{r|}{\bf{59}}  & \bf{70}  \\ 
\hline

\end{tabular}
\label{tbl-rag-ft-datab}
\end{center}
\end{table}

\begin{table}[H]
\caption{Baseline, RAG, and FT for STM.}
\begin{center}
  \scriptsize
\begin{tabular}{|l|rr|rr|rr|}
\hline
  \textbf{STM}  & \multicolumn{2}{c|}{\textbf{Granite-8b}} & \multicolumn{2}{c|}{\textbf{Granite-8b}}  & \multicolumn{2}{c|}{ \bf{Granite-8b} } \\
  \textbf{Tests} & \multicolumn{2}{c|}{\textbf{Base}} & \multicolumn{2}{c|}{\textbf{RAG}}  & \multicolumn{2}{c|}{ \bf{FT} } \\\cline{2-7} 
                       & \multicolumn{1}{r|}{\textbf{Opt}}  & \textbf{Full}& \multicolumn{1}{r|}{\textbf{Opt}}  & \textbf{Full}   & \multicolumn{1}{r|}{\textbf{Opt}}  & \textbf{Full} \\ \hline
else\_body         & \multicolumn{1}{r|}{60}  & 7,296  & \multicolumn{1}{r|}{\bf{58}}  & 7,868   & \multicolumn{1}{r|}{78}   & \textbf{157}\\ 
\hline
for\_body         & \multicolumn{1}{r|}{\bf{142}}  &   7,808 & \multicolumn{1}{r|}{146}  &   7,374  & \multicolumn{1}{r|}{153}   & \bf{159}  \\
\hline
func\_body        & \multicolumn{1}{r|}{343}  & 5,210 & \multicolumn{1}{r|}{226}  &   6,152    & \multicolumn{1}{r|}{\bf{224}}   & \bf{266} \\ 
\hline
if\_body         & \multicolumn{1}{r|}{\bf{132}}  &  10,173  & \multicolumn{1}{r|}{226}  &  6,152   & \multicolumn{1}{r|}{144}   & \bf{197} \\ 
\hline
logging            & \multicolumn{1}{r|}{40}   &  8,046  & \multicolumn{1}{r|}{39}   &    7,656   & \multicolumn{1}{r|}{\bf{29}}   & \bf{30} \\
\hline
func\_call       & \multicolumn{1}{r|}{40}   &   7,288    &\multicolumn{1}{r|}{37}   &  7,912   & \multicolumn{1}{r|}{\bf{21}}   & \bf{21}\\
\hline
\hline
  \textbf{STM}  & \multicolumn{2}{c|}{\textbf{Llama-8b}} & \multicolumn{2}{c|}{\textbf{Llama-8b}}  & \multicolumn{2}{c|}{\textbf{Llama-8b}} \\ 
   \textbf{Tests}  & \multicolumn{2}{c|}{\textbf{Base}} & \multicolumn{2}{c|}{\textbf{RAG}}  & \multicolumn{2}{c|}{\textbf{FT}} \\\cline{2-7} 
                       &  \multicolumn{1}{r|}{\textbf{Opt}}  & \textbf{Full} &  \multicolumn{1}{r|}{\textbf{Opt}}  & \textbf{Full}  & \multicolumn{1}{r|}{\textbf{Opt}}  & \textbf{Full} \\ \hline

else\_body         & \multicolumn{1}{r|}{118}  &  11,174  & \multicolumn{1}{r|}{92} & 12,010   & \multicolumn{1}{r|}{\textbf{42}}   & \bf{393} \\ 
\hline
for\_body          & \multicolumn{1}{r|}{987}  & 11,779 & \multicolumn{1}{r|}{169} & 12,918  & \multicolumn{1}{r|}{\bf{163}}   & \bf{186} \\
\hline
func\_body         & \multicolumn{1}{r|}{1,210}  &  11,295 & \multicolumn{1}{r|}{264} & 12,300   & \multicolumn{1}{r|}{\bf{195}}   & \bf{225} \\ 
\hline
if\_body            & \multicolumn{1}{r|}{437}  &  10,678  & \multicolumn{1}{r|}{148} & 11,438 & \multicolumn{1}{r|}{\bf{133}}   & \bf{183} \\ 
\hline
logging           & \multicolumn{1}{r|}{\bf{39}}  & 13,316 & \multicolumn{1}{r|}{61}   & 12,395 & \multicolumn{1}{r|}{\bf{44}}   & 44  \\
\hline
func\_call         & \multicolumn{1}{r|}{56}  &   10,914 & \multicolumn{1}{r|}{50} & 10,825  & \multicolumn{1}{r|}{\bf{43}}   & \bf{52}  \\
\hline
\end{tabular}
\label{tbl-rag-ft-stm}
\end{center}
\end{table}

The repositories consistency of style and the application of best practices within a category has a strong influence on how well a model can guess successfully without requiring a prompt. For the vast majority of the 24 combinations the customized models show improvements in opt, which are for some combinations very large. RAG customized models failed consistently to achieve conciseness. Categories with more consistent patterns as function body, logging and function calls enabled the models to adapt better than categories with more variability. For the full Levenshtein distance the FT customized models produced the best result in all combinations.

\begin{table}[h]
\caption{Baseline and FT Model Results for CCEVal and RepoBench benchmarks. }
\begin{center}
\scriptsize
\begin{tabular}{|l|rrrr|rr|}
\hline
 & \multicolumn{6}{c|}{\bf{Granite}} \\ \cline{2-7}
                \bf{Benchmark}  & \multicolumn{2}{c|}{\bf{Base}}  & \multicolumn{2}{c|}{\bf{FT 3ep}} & \multicolumn{2}{c|}{\bf{FT 10ep}}\\ \cline{2-7} 
                  & \multicolumn{1}{c|}{\bf{Opt}} & \multicolumn{1}{c|}{\bf{Full}} & \multicolumn{1}{c|}{\bf{Opt}} & \bf{Full} & \multicolumn{1}{c|}{\bf{Opt}} & \bf{Full} \\ \hline

CCEeval           & \multicolumn{1}{r|}{29} & \multicolumn{1}{r|}{3,125}  & \multicolumn{1}{r|}{32}   & 34  & \multicolumn{1}{r|}{29}   & 30  \\ \hline
RBench        & \multicolumn{1}{r|}{20}   & \multicolumn{1}{r|}{4,138}  & \multicolumn{1}{r|}{30}   & 36  & \multicolumn{1}{r|}{9}   &  9 \\ \hline
\hline
  & \multicolumn{6}{c|}{\bf{Llama}} \\ \cline{2-7} 
                \bf{Benchmark}  & \multicolumn{2}{c|}{\bf{Base}}  & \multicolumn{2}{c|}{\bf{FT 3ep}} & \multicolumn{2}{c|}{\bf{FT 10ep}}
                  \\ \hline

CCEeval           & \multicolumn{1}{r|}{51} & \multicolumn{1}{r|}{8,445}   & \multicolumn{1}{r|}{41}   & 42   & \multicolumn{1}{r|}{40} & 41 \\ \hline
RBench       & \multicolumn{1}{r|}{74}  & \multicolumn{1}{r|}{9,510}     & \multicolumn{1}{r|}{25}   & 25  & \multicolumn{1}{r|}{21} & 21 \\ \hline
\end{tabular}
\end{center}

\label{tbl_cceval_repobench}
\end{table}

Table \ref{tbl_cceval_repobench} shows results for FT customization for the public benchmarks, confirming the potential for improvement across different data sets and base models. The progression of the distance measurements with training illustrates the importance of good convergence and sufficient training epochs for 
the 'opt' distance.

\section{Conclusions}
\label{sec:conclu}

We presented our repository-level automated data preparation and model customization methodology on enterprise repository data for code completion. 
Our results demonstrate how the strategy of ingesting and creating training and testing samples based on semantic scopes can help a model understand the implicit usage patterns from a repository and predict more accurate and concise code that aligns with the ground truth without requiring human labor to generate training or RAG data.
Furthermore we discussed methodology aspects as properties of measurements for the code completion scenario, numerical instabilities and inaccuracies in nearest neighbor searches and options to reduce their impact on convergence and prediction quality. 

User testimony and quantitative evaluation on large private repositories with two base models and multiple categories of developer guided tests shows that customizing the models via supervised FT achieves the best performance compared to off-the-shelf models as well as off-the-shelf models combined with RAG. It also establishes that customization of smaller models, with much lower response latencies, can provide a superior user experience even compared to out-of-the box models that are 10 times larger and hence have correspondingly higher response latencies. After effort-to-value ratio and quality and conciseness, latency was an important criterion for user experience. 

\section{Future Work}
In future work we will expand the implementation of the semantic scope-based data ingestion and model customization to other enterprise repositories, including other coding tasks. Also consider post-training via Reinforcement Learning from Human Feedback (RLHF), and analyze the effect of customized models in Agentic frameworks. 

\section*{Acknowledgements}
We thank Xuan Liu and Nicholas Fuller for supporting our work. We thank Aslam Nomani and team for their technical insight and feedback.

\bibliographystyle{plain}
\bibliography{conference_biblio}

\end{document}